\documentclass[12pt]{iopart}

\usepackage{graphicx}

\newcommand{\grsim}{\mbox{\raisebox{-0.6ex}{$\stackrel{>}{\sim}$}}\:}

\begin{document}

\title[Hydrodynamic afterburner for the CGC at RHIC]{Hydrodynamic afterburner for the CGC at RHIC}

\author{Tetsufumi Hirano
 \footnote[1]{Present address: Department of Physics, Columbia University, New York, NY 10027, USA} and 
Yasushi Nara
 \footnote[2]{Present address: Institute f\"ur Theoretische Physik,
J.~W.~Goethe Universit\"at, D-60054 Frankfurt, Germany}
}

\address{\dag RIKEN BNL Research Center, Brookhaven National Laboratory,
Upton, NY 11973, USA
}

\address{\ddag Department of Physics, University of Arizona, Tucson, Arizona 85721, USA}

\begin{abstract}
Firstly, we give a short review
about the hydrodynamic model and its application
to the elliptic flow phenomena 
in relativistic heavy ion collisions.
Secondly, we show the first approach
to construct a unified model for the description of
the dynamics in relativistic heavy ion collisions.
\end{abstract}




\section{A Short Review on Elliptic Flow from Hydrodynamic
models}

First data reported by the STAR Collaboration at 
RHIC \cite{STAR:v2} has a significant meaning that
the observed large magnitude of elliptic flow
for charged hadrons
is consistent with hydrodynamic predictions
\cite{KHHH}.
This suggests that
large pressure possibly in the partonic phase
is built at the early stage
($\tau \sim 0.6$ fm/$c$)
in Au+Au collisions at $\sqrt{s_{NN}} =$ 130 and 200 GeV.
This situation at RHIC is in contrast to
that at lower energies such as AGS or SPS
where hydrodynamics always overpredicts
the data \cite{NA49:v12}.
Moreover, this also suggests that the effects of the viscosity
in the QGP phase is remarkably small and that the QGP is 
almost a perfect fluid \cite{S}.
Hadronic transport models
are very good to describe experimental data
at lower energies, while they fail to reproduce
such large values of
elliptic flow parameter at RHIC (see, e.g., Ref.~\cite{BS}).
So the importance of hydrodynamics is rising
in heavy ion physics.
After the first STAR data were published \cite{STAR:v2}, 
other groups at RHIC have also obtained the data
concerning with flow phenomena \cite{Retiere}.
To understand
these experimental data,
hydrodynamic analyses are also performed
extensively \cite{QGP3:hydro:Pasi,QGP3:hydro:Peter}.
In this short review, we highlight several
results on elliptic flow
from hydrodynamic calculations.

\subsection{Why elliptic flow?}

Elliptic flow is very sensitive to the degree of secondary
interactions \cite{O}.
The indicator of momentum anisotropy
is the second harmonic coefficient
of azimuthal distributions \cite{PV}
\begin{eqnarray}
v_2(p_T, y) & = & \frac{\int d\phi \cos(2\phi)\frac{dN}{p_Tdp_T dy d\phi}}{\int d\phi \frac{dN}{p_T dp_T dy d\phi}} = \langle \cos(2\phi) \rangle.
\label{eq:v2}
\end{eqnarray}
For recent progress of higher harmonics, see Refs.~\cite{P,AP}.
The spatial anisotropy
\begin{eqnarray}
\varepsilon = \frac{\langle y^2 - x^2 \rangle_s}{\langle x^2+y^2 \rangle_s},
\quad \langle \cdots \rangle_s = \frac{\int \cdots e(\tau, x)d^3x}{\int e(\tau, x)d^3x}
\end{eqnarray}
at the initial time is a seed of the momentum anisotropy.
Within the hydrodynamic picture, pressure gradient along
the $x$-axis is larger than the $y$-axis in non-central collisions.
Here the direction of $x$-axis is parallel to the impact parameter vector.
This is the driving force for generating $v_2$.
The response of the system can be defined by $v_2/\varepsilon$.
As will be discussed later, the hydrodynamic response
is almost constant $v_2/\varepsilon \sim 0.2$-0.24 at the SPS
and RHIC energies \cite{KSH}.

\begin{figure}[t]
\begin{center}
\includegraphics[width=0.45\textwidth]{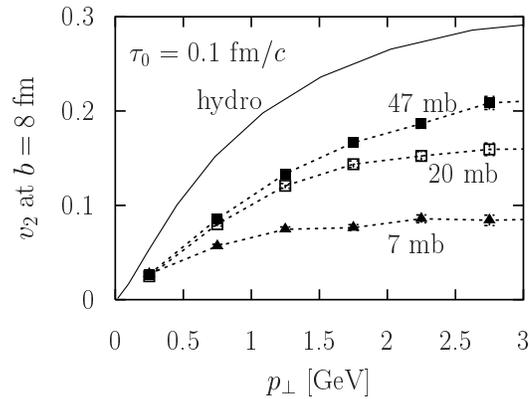}
\end{center}
\caption{
Comparison of the hydrodynamic result with
the results obtained by a partonic transport model.
Figure taken from Ref.~\cite{DenesPasi}.
}
\label{fig:hydroBoltzmann}
\end{figure}

In ideal hydrodynamics, the mean free path among
particles is assumed to
be zero. So the hydrodynamic prediction of $v_2$
should be maximum among transport theories.
We naively expect that the ideal hydrodynamics corresponds to
the infinite limit of the cross sections
among particles
in collision terms of the Boltzmann equation.
In Ref.~\cite{DenesPasi}, Molnar and Huovinen compare
$v_2(p_T)$ from hydrodynamics with the one from Boltzmann equation.
Special attention is paid that the same initial conditions
and thermodynamic properties are used in comparison.
Figure \ref{fig:hydroBoltzmann} shows
the transverse momentum dependence of $v_2$ for a hydrodynamic model
and a parton cascade model.
$v_2(p_T)$ from Boltzmann simulations is approaching
to the hydrodynamic result
with increasing the cross section among partons.
However, even for large cross section (47 mb),
the result from the Boltzmann simulation
is still $\sim$ 30\% smaller than the hydrodynamic result.
This clearly shows
the ideal hydrodynamics describes the strongly interacting
matter.

\subsection{Basics of Ideal Hydrodynamics}

Here we concentrate our discussions on the ideal hydrodynamic model.
For those who have interest in the viscous hydrodynamics,
see, e.g., Ref.~\cite{M}.

Hydrodynamic equations are nothing but the
energy-momentum conservations
$\partial_\mu T^{\mu\nu}=0$.
In the ideal hydrodynamics, the energy-momentum tensor 
becomes $T^{\mu\nu} = (e+P) u^\mu u^\nu - Pg^{\mu\nu}$,
where $e$ is the energy density, $P$ is the pressure,
and $u^\mu$ is the four fluid velocity.
When there are conserved quantities 
such as the baryon number or the number of chemically frozen
hadrons,
one needs to solve
the continuity equations
$\partial_\mu n_i^\mu = 0$ together with the hydrodynamic equations.
In order to close the system of partial differential
equations, the equation of state (EoS)
$P(e,n_i)$ is needed.
It is the EoS that
governs the dynamics of the system in the ideal hydrodynamics.
The naive applicability conditions of ideal
hydrodynamics are that the mean free path among the
particles is much smaller than the typical size
of the system and that the system keeps local thermal equilibrium
during expansion.
From these conditions, one cannot use hydrodynamics
for initial collisions, final free streaming
and high $p_T$ ($\grsim$ 2 GeV/$c$) particles.
The quantitative criteria are more complicated
in general since these come from
dynamical aspects of the space-time evolution.

One usually assumes the initial hydrodynamic
fields at initial (or equilibrated) time
around 1 fm/$c$.
One needs an interface between
the pre-thermalisation stage and the hydrodynamic
stage at the initial time.
On the other hand,
the system eventually breaks up and the assumption
of the thermalisation is no longer valid
at the later stage.
This means a prescription of freezeout
is needed in the hydrodynamic model when one evaluates
the particle spectra to be observed by the detectors.
Then one needs another interface
between the hydrodynamic stage and
the free streaming stage.
In what follows, we discuss particularly
equation of state, initial condition
and freezeout prescription used in the literature.

\subsubsection{Equation of State}

The main ingredient of the hydrodynamic model is
the equation of state (EoS) for thermalised
matter produced in heavy ion collisions.
Ideally, one uses the EoS taken from
the first principle
calculations of QCD, namely,
lattice QCD simulations \cite{Karsch}.
More practically,
one can use the resonance gas model for the hadron phase and
the massless free parton gas for the QGP phase.
By matching these two EoS's at the critical temperature,
one obtains the first order phase transition model
with a latent heat $\sim$1 GeV/fm$^3$.
In the mixed phase in this model, the sound velocity
$c_s^2 = \frac{\partial P}{\partial e}$
is vanishing.
Recent lattice QCD simulations
tell us that the phase transition
seems to be crossover in vanishing baryonic chemical
potential.
Although the discontinuity of the thermodynamic
variables do not exist in the crossover
phase transition,
it should be emphasized that
the energy density and the entropy density increase
more rapidly than the pressure
in the vicinity of
the phase transition region $\Delta T \sim 0.1 T_c$.
This also leads to very small sound velocity
near the phase transition region.
Therefore it is very hard to find flow observables which
distinguish the crossover phase transition with
a rapid change of the thermodynamic variables
from the first order phase transition.

The statistical model and the blast wave model
tell us that chemical freezeout temperature $T^{\mathrm{ch}}$
is larger than thermal freezeout temperature $T^{\mathrm{ch}}$
at the RHIC energies (see, e.g., discussions in Ref.~\cite{SH}).
Chemical freezeout temperature obtained from the statistical model
analysis is very close to the (pseudo)critical temperature
of the QCD phase transition.
This means that the hadron phase is almost chemically
frozen due to strong expansion.
For chemically frozen hadronic gas,
one introduces
chemical potential for each hadron \cite{bebie}.
This modifies the chemical composition in the hadron
phase compared with the chemical equilibrium
hadronic gas and, consequently,
changes the space-time evolution of
temperature field \cite{HiranoTsuda}.

\subsubsection{Initial Condition}

Once initial conditions are assigned, one can numerically simulate
the space-time evolution
of thermalised matter 
which is governed by hydrodynamic equations.
Usually, initial conditions are parametrised
based on some physical assumptions.
Transverse profile of the energy/entropy
density is assumed to be proportional to
the local number of participants $\rho_{\mathrm{part}}$,
the local number of binary collisions $\rho_{\mathrm{coll}}$
or linear combination of them.
Initial transverse flow is usually taken to be vanishing.
In either full three dimensional (3D)
hydrodynamic simulations or hydrodynamic
simulations of longitudinal
expansion with cylindrically
symmetric geometry,
one needs to parametrise also longitudinal profile
of initial conditions
for energy density, baryon density
and four fluid velocity.
The initial longitudinal shape is chosen so as to
reproduce the final (pseudo)rapidity
distribution of hadrons.
Unfortunately, the shape
is not uniquely determined.
Two completely different initial conditions
can end up almost similar rapidity distribution.
See, e.g., Ref.~\cite{Huovinen:1998tq}.

On the other hand, one can introduce
model calculations
to obtain the initial condition
of hydrodynamic simulations.
Event generators can be used
to obtain the energy density distribution
at the initial time.
Recently, the SPheRIO group
employs an event generator NeXus
and takes an initial condition
from this model
in the event-by-event basis \cite{OAHK}.
The resultant energy density
distribution in the transverse
plane has highly bumpy structures \cite{OAHK,GRZ}.
Smooth initial conditions 
used in the
conventional hydrodynamic simulations
are no longer expected in one event.
Another important example
which is relevant
at very high collision energies
is an initial condition
taken from the Colour Glass Condensate
picture \cite{HiranoNara6}.
This will be discussed in details
in Sec.~2.

\subsubsection{Freezeout}

Conventional prescription to obtain the invariant momentum
spectra
from the hydrodynamic simulations
is to employ the so-called Cooper-Frye formula
\cite{CF}
\begin{eqnarray}
\label{eq:CF}
E\frac{dN_i}{d^3p} = \frac{d_i}{(2\pi)^3} \int_{\Sigma(x)}
\frac{p^\mu d\sigma_\mu(x)}
{\exp[(p^\nu u_\nu(x)-\mu_i(x))/T(x)]\pm 1}
\end{eqnarray}
where $p^\mu = (E,\mathbf{p})$ and $d_i$ are, respectively,
a four momentum and the degree of freedom of a particle
under consideration.
Integral is performed on the freezeout hypersurface $\Sigma$.
Here $d\sigma$ is an element of freezeout hypersurface on $\Sigma$.
One usually assumes $\Sigma = \Sigma(T=T^{\mathrm{th}}\enskip
 \mathrm{or} \enskip e = e^{\mathrm{th}})$, where $T^{\mathrm{th}}$ ($e^{\mathrm{th}}$) is
thermal freezeout temperature (energy density).
$T^{\mathrm{th}}$ or $e^{\mathrm{th}}$
is an adjustable parameter for
reproduction of the slope of $p_T$ spectrum.
Information on the hydrodynamic simulations
is taken through $d\sigma$, $u^\mu$, $T$ and $\mu_i$ at a space-time
point $x$.
Chemical potential for each hadron $\mu_i$
is not vanishing in general even in vanishing baryon density
due to $T^{\mathrm{ch}}>T^{\mathrm{th}}$.
Although the chemical potential is important to reproduce
the particle ratios as well as the shape of
the particle spectra, this is often neglected
for simplicity in the literature.

The physical picture described by the Cooper-Frye
formula
is sometimes called ``sudden freezeout" since
the mean free path $\lambda$ 
is suddenly changed from zero to infinity
through a thin freezeout hypersurface.
Instead of using this,
one can use a hadronic cascade model
to describe the space-time
evolution of hadrons \cite{BD,TLS}.
The mean free path among hadrons
is finite and depends on hadronic species.
Hence, 
one can describe a continuous
freezeout picture
through hadronic transport models.
Note that continuous
particle emission can be considered
within the hydrodynamics \cite{Grassi:1994nf}.
Adoption of hadronic transport models
after hydrodynamic evolution of the QGP liquid
could refine the dynamical modeling
of relativistic heavy ion collisions.
However, it is not so easy
to connect them in a systematic and proper way.
One has to use Eq.~(\ref{eq:CF}) again
for the boundary between the QGP liquid
and the hadron gas. However,
Eq.~(\ref{eq:CF}) describes not only
out-going particles $p^\mu d\sigma_\mu>0$ but also
in-coming particles $p^\mu d\sigma_\mu<0$.
If one uses hydrodynamic simulations
as inputs for hadronic cascade models,
one must discard the in-coming particles,
which violates the energy-momentum conservation.
In order to remedy this problem,
one need to solve hydrodynamic equations
\textit{together with} the appropriate kinetic equations
\cite{Bugaev}.

\subsection{Hydrodynamic Results for $v_2$}

We show
comparison of the hydrodynamic results with
the experimental data for
$v_2(p_T)$, $v_2(\eta)$ at the RHIC energy 
and the excitation function of $v_2$.

\subsubsection{$v_2(p_T)$}
\label{sec:v2pt}

\begin{figure}[t]
\begin{center}
\includegraphics[width=0.4\textwidth]{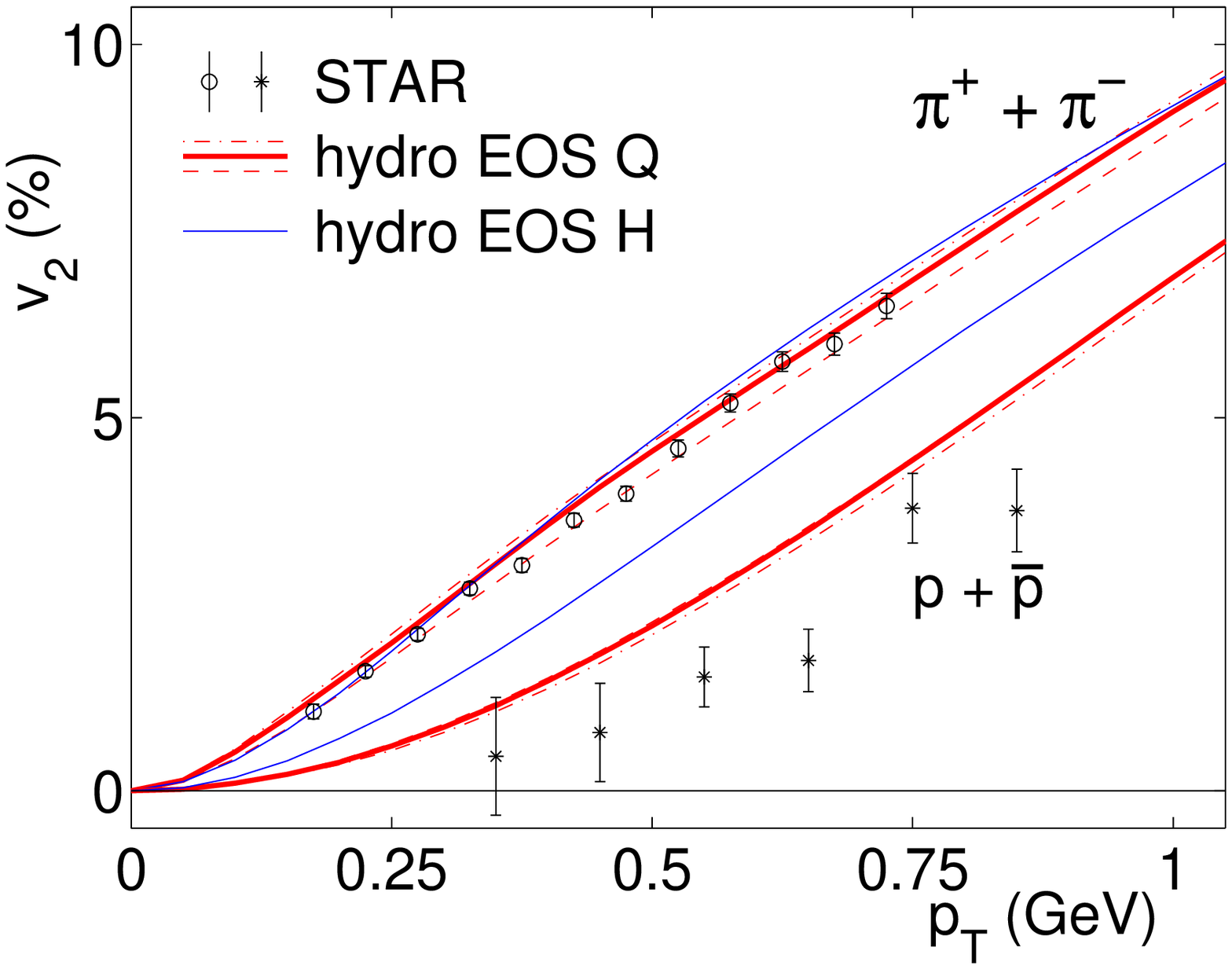}
\includegraphics[width=0.5\textwidth]{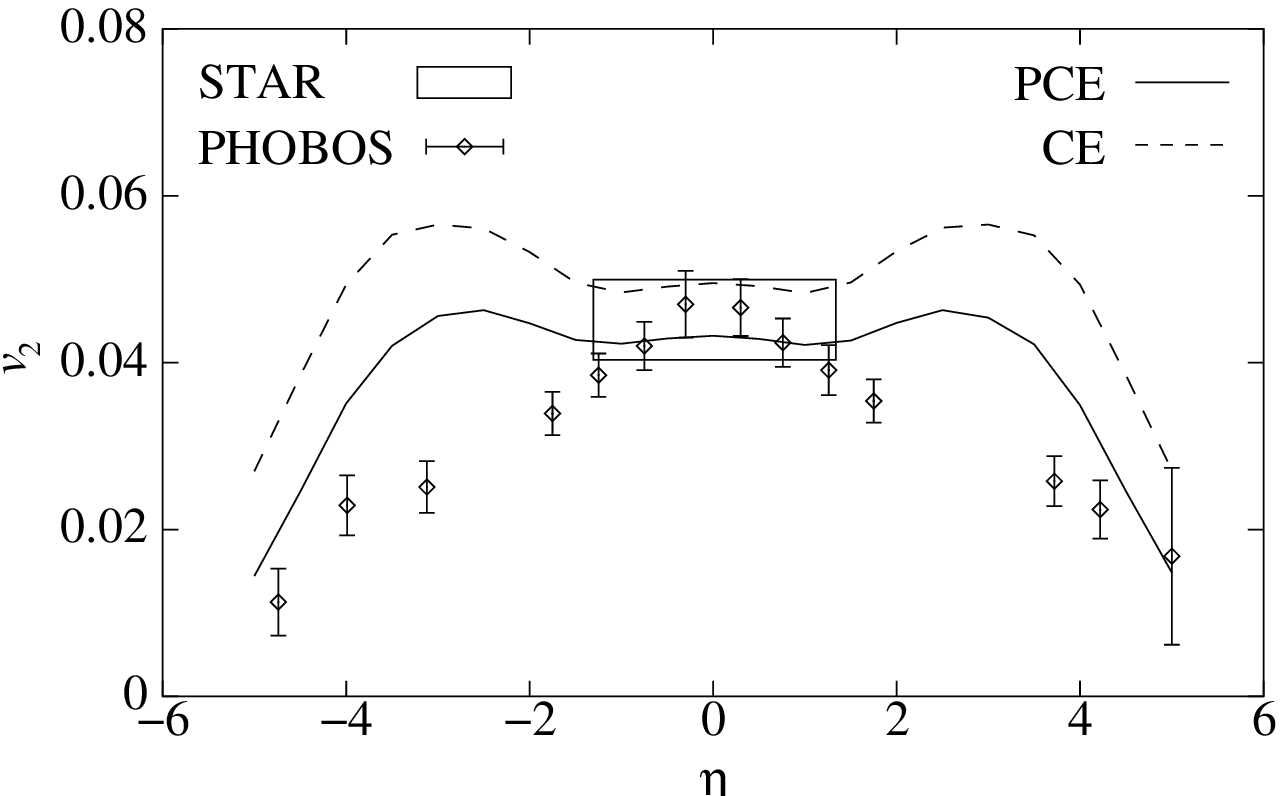}
\end{center}
\caption{
(left)
Transverse momentum dependence of $v_2$ for pions and (anti)protons
\cite{QGP3:hydro:Peter} are compared with the STAR data \cite{STAR:v2id}
in minimum bias Au+Au collisions at $\sqrt{s_{NN}}=130$ GeV.
EoS Q stands for the equation of state with first order
phase transition between the massless free parton gas and
the resonance gas. On the other hand, EoS H stands
for the equation of state for the resonance gas.
(right)
$v_2$ for charged hadrons
as a function of pseudorapidity $\eta$
in Au+Au collisions at $\sqrt{s_{NN}}=130$ GeV \cite{STAR:v2,PHOBOS:v2eta}.
PCE means the EoS of partial chemical equilibrium,
whereas CE means the EoS of chemical equilibrium.
Figure is taken from Ref.~\cite{HiranoTsuda}.
}
\label{fig:v2pt}
\end{figure}

With a help of assuming the Bjorken flow \cite{BJ}
for the longitudinal direction,
one can solve the hydrodynamic equations
only in the transverse plane at midrapidity.
Systematic studies based on this (2+1)-dimensional hydrodynamic
model are performed in Ref.~\cite{KHHH}.
For the EoS, complete chemical equilibrium is assumed for both
the QGP phase and the hadron phase.
$p_T$ dependences of $v_2$ for pions and protons 
from this model \cite{QGP3:hydro:Peter} are
compared with the STAR data \cite{STAR:v2id}
in Fig.~\ref{fig:v2pt} (left).
By employing the EoS with phase transition,
the hydrodynamic model correctly
reproduces $v_2(p_T)$ and its mass-splitting behavior
below $p_T=1$ GeV/$c$.
On the other hand, $v_2(p_T)$ for (anti)protons
from the resonance gas model
does not agree with the data.
Although the reason for the difference
of the result between these two EoS models
is not so clear,
the experimental data favors
the QGP EoS.
Due to the assumption of chemical equilibrium
in the hadron phase, 
this model does not reproduce particle ratio
and spectra simultaneously.
It is of importance to study whether the agreement
with the experimental data still holds
even when the assumption of chemical equilibrium
in the hadron phase
is abandoned \cite{HiranoTsuda}.


\subsubsection{$v_2(\eta)$}
\label{sec:v2eta}
One needs a full 3D
hydrodynamic simulation 
to obtain the rapidity dependence of $v_2$
since either
cylindrical symmetry or Bjorken flow \cite{BJ}
keeps us from obtaining this quantity.
First analysis of $v_2(\eta)$ at RHIC based on
the full 3D hydrodynamic model
is performed in Ref.~\cite{Hirano}. 
After that, the effect of early chemical freezeout
on the elliptic flow is also investigated \cite{HiranoTsuda}.
Figure \ref{fig:v2pt} (right) shows $v_2(\eta)$ for charged hadrons
in Au+Au collisions at $\sqrt{s_{NN}}=130$ GeV \cite{STAR:v2,PHOBOS:v2eta}.
Here the initial condition for the energy density is so
chosen as to reproduce the pseudorapidity distribution
of charged hadrons.
Hydrodynamic results are consistent with
the experimental data only near the midrapidity
while hydrodynamics overpredicts
the data in the forward/backward rapidity regions.
Multiplicity is not so large in the forward/backward
rapidity regions, so equilibration of the system
tends to be spoiled.

\begin{figure}[t]
\begin{center}
\includegraphics[width=0.35\textwidth]{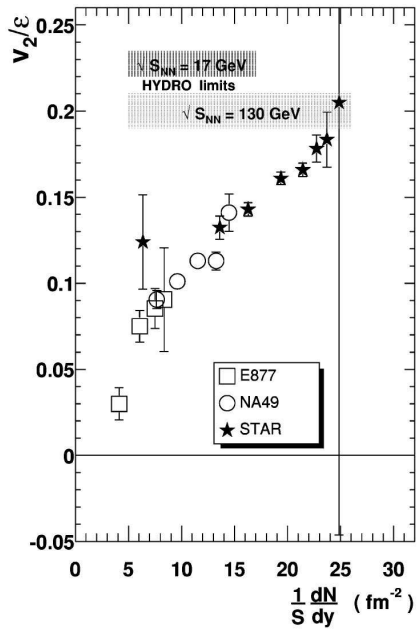}
\end{center}
\caption{
Excitation function of $v_2/\varepsilon$.
Figure taken from Ref.~\cite{STAR:scaledv2}.
}
\label{fig:v2overeps}
\end{figure}

\subsubsection{Excitation function}

Figure \ref{fig:v2overeps} shows the excitation
function of $v_2$ compiled
by the STAR Collaboration \cite{STAR:scaledv2}.
Hydrodynamic results presented in this
figure are based on the same model
discussed in Sec.~\ref{sec:v2pt}.
Data points continuously
increase with the unit rapidity density
at the AGS, SPS and RHIC energies.
However, the hydrodynamic response
is almost flat or slightly decreases
with the multiplicity.
The data points seem to reach the ``hydrodynamic limit"
for the first time at the RHIC energy.
It will be interesting to see the same plots including the upcoming LHC
results: If the system produced at LHC obeys the hydrodynamic picture,
this experimental ``curve" will bend at $(1/S)dN/dy\sim 25$-30.

The deviation between the hydrodynamic results and
the data plots below $(1/S)dN/dy = 25$
reminds us the pseudorapidity dependence of elliptic flow
in Fig.~\ref{fig:v2pt} (right). 
The deviation might come from a common origin \cite{Heinzqm2004}:
the small multiplicity both in forward rapidity region
at the RHIC energy
and at midrapidity at the lower collision energies
could cause the partial thermalisation.
In the low density limit, $v_2$ is actually
proportional to the number density \cite{Heiselberg}.
The shape of $v_2(\eta)$ data in forward rapidity region
looks similar to that of
the pseudorapidity distributions \cite{Steinberg}.
Similarly, data plots of excitation function increase
almost linearly with the particle density.
These results suggest that thermalisation
is not achieved completely in forward rapidity region
at the RHIC energy and at midrapidity at the SPS energies.

\begin{figure}[t]
\begin{center}
\includegraphics[width=0.45\textwidth]{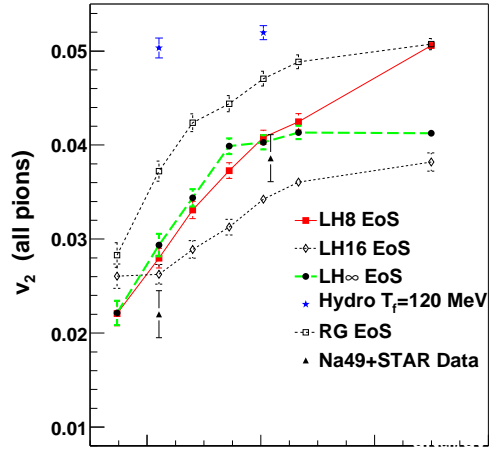}
\end{center}
\caption{
Excitation function from the hydro+cascade model.
Figure taken from Ref.~\cite{TLS}.
}
\label{fig:H2H}
\end{figure}

Figure \ref{fig:H2H} shows
the excitation function based on the hydro+cascade
model \cite{TLS}
with Bjorken longitudinal flow \cite{BJ}.
In the hadron phase, a hadronic event generator
RQMD is employed in these calculations.
Contrary to the conventional hydrodynamic models,
freezeout processes are automatically
described by this model without any adjustable parameters.
The excitation function in the case of 
the latent heat $\sim 0.8$ GeV/fm$^3$
linearly increases
with the multiplicity, which is consistent with
the experimental data.
It should be noted that $v_2(p_T)$, its mass dependences
and particle spectra/ratio at midrapidity
are also reproduced by this hybrid model \cite{TLS}.

\subsection{Summary and Outlook}

The most successful approach based on the hydrodynamic
model to elliptic flow
in relativistic heavy ion collisions is the hybrid one in which
the QGP phase is described by
the ideal hydrodynamics
while the hadron phase is described by a
hadronic cascade.
The Bjorken scaling solution \cite{BJ} is assumed 
in the current hydro+cascade model.
This means that current hybrid models
are available only near midrapidity.
Therefore it is desired to develop a model
in which a full 3D hydrodynamic
simulation is combined with
a hadronic cascade model.
From agreement of excitation function
between the hybrid model and the experimental
data at midrapidity, the 3D hybrid model is expected to reproduce
the pseudorapidity dependence of elliptic
flow.
It should be emphasised again
that the hybrid model has its own problem
on the violation of energy momentum conservations
between the QGP liquid
and the hadron gas.
A simulation which incorporates a proper
treatment at the boundary between the QGP phase and
the hadron phase is now an open problem.

\section{Hydrodynamic Model with a CGC Initial Condition}
\label{sec:2}

\subsection{Toward a Unified Description of Relativistic Heavy Ion Collisions}

As already discussed in the previous section, one of the important
findings at RHIC is that the hydrodynamic approach to the
description of elliptic flow works remarkably well
when early thermalisation time
and the QGP EoS are assumed \cite{QGP3:hydro:Pasi,QGP3:hydro:Peter}.
Hydrodynamics should work only in the situation that the mean free
path $\lambda = 1/\sigma \rho$ is much smaller than
the typical size of the system $L$. This suggests that
the particle density $\rho$ becomes very large at RHIC.

In addition to this, there are many important findings in
heavy ion collisions at the RHIC energy.
One of them is jet quenching \cite{QGP3:highpt}, i.e., 
 suppression of high $p_T$ hadrons
in the single particle spectra and disappearance of away-side
peak in the di-hadron spectra. These phenomena are not found
in $dAu$ collisions \cite{PHOBOS:dA,PHENIX:dA,STAR:dA,BRAHMS:dA}:
the yield of high $p_T$ hadrons
roughly scales with the number of binary collisions $N_{\mathrm{coll}}$
and the away-side peak appears
in the di-hadron spectra.
So one can conclude that high $p_T$ partons produced
in the initial hard scattering
interact with the highly \textit{dense} medium
in the final state
and that these partons lose their energies during
traversing medium.

So the current RHIC data strongly
suggest that the parton density
created in heavy ion collisions at RHIC
is dense enough to cause both large elliptic flow
of bulk matter
and large suppression of high-$p_T$ hadrons.
What is an origin of this dense matter at RHIC?
The bulk particle production in high energy hadronic/nuclear
collisions  is dominated by the small $x$
modes in the nuclear wave function, where $x$ is a momentum
fraction of the incident particles.
Hence, an origin of the large density could be traced to the initial
parton density at small $x$ inside the ultra-relativistic nuclei
 before collisions.
It is well known that the gluon density increases rapidly 
with decreasing $x$ until gluons begin to overlap
in the phase space
where nonlinear interaction becomes important \cite{GLR83}.
These gluons eventually form
another type of extremely dense matter,
the Colour Glass Condensate (CGC)~\cite{QGP3:cgc}.
This phenomenon is characterised by a ``saturation scale"
$Q_s^2$ which is identified with the gluon rapidity
density per unit transverse area.
When $Q_s$ is sufficiently larger
than the typical hadronic scale
 ($Q_s \gg \Lambda_{\mathrm{QCD}}$),
the strong coupling constant
 becomes weak ($\alpha_s(Q_s)\ll 1$).
Moreover, the gluon occupation number in the wave function
becomes huge, $\sim 1/\alpha_s(Q_s) \gg 1$.
Therefore the CGC
can be studied by an weak coupling classical method
also known as McLerran-Venugopalan (MV) model~\cite{MV}.
The calculations based on the CGC picture have been compared to
 various RHIC data.
Remarkably, the CGC results on the global observables, e.g.,
the centrality, rapidity and energy dependences of charged
hadron multiplicities agree with the RHIC data under the assumption
of parton-hadron duality~\cite{KLN}.

In relativistic heavy ion collisions, the CGC 
could be a seed of the QGP \cite{GyulassyMcLerran}.
So hydrodynamic initial conditions
can be calculated from the two CGC collisions.
One needs a non-equilibrium model
which describes equilibration from the produced gluon
distribution to a local thermal distribution.
This is, however, beyond the present work.
In this work, the rapidity, transverse position
and centrality dependences of the number of
produced gluons are identified with
the hydrodynamic initial condition of a thermalised QGP.

Some of the problems which are inherent
in a particular approach can be removed.
For instance, 
one conventionally
parametrises initial conditions in the hydrodynamic simulations
shortly after a collision of two nuclei by hand.
Therefore it is desired to take an initial condition which
is obtained by a reliable theory.
On the other hand, most of the calculations based on
the CGC do not 
include final state interactions.
According to estimation of initial parton productions
 from a classical lattice Yang-Mills
simulation based on the CGC~\cite{KNV3,Lappi},
the transverse energy per particle
is roughly $E_T/N \sim 1.5$ GeV,
while the experimental data yields $E_T/N \sim 0.6$ GeV \cite{PHENIX:meanet}.
Moreover, elliptic flow from a classical Yang-Mills simulation is
inconsistent with RHIC data~\cite{KNV2}.
Those discrepancies, which are mainly due to the lack
of collective effects, will be removed by introducing further time
evolution through, e.g., hydrodynamics.
For the calculation of parton energy loss, 
one needs time evolution of parton density.
Bjorken expansion \cite{BJ} is assumed 
for the evolution of parton density
in almost all work except Refs.~\cite{GVWH,HiranoNara}.

We have already developed a unified model
(CGC+Hydro+Jet model)
in which hydrodynamic evolution, CGC initial conditions,
and the production and propagation of high $p_T$
partons are combined \cite{HiranoNara6}.
Only the soft sector, i.e., the production of gluons 
in CGC collisions and
the hydrodynamic afterburner for the CGC initial condition
is discussed in this paper.
For further details of the model
and the results for high $p_T$
regions such as
centrality dependences of the nuclear modification factors
and the azimuthal correlation functions,
see Ref.~\cite{HiranoNara6}.

\subsection{Gluon Production from a CGC picture}

We employ the $k_T$ factorisation formula
for the production of gluons from CGC collisions
 along
the line of work by Kharzeev, Levin and Nardi (KLN)~\cite{KLN}.
The number of produced gluons
in the $k_T$-factorisation formula
 is given by~\cite{GLR83,GLR81,LL94,Szczurek:2003fu}
\begin{eqnarray}
\frac{dN_g}{d^2x_{\perp}dy}&=&
   \frac{4\pi^2N_c}{N_c^2-1} \int\frac{d^2p_T}{p^2_T}
   \int d^2k_T \alpha_s(Q^2) \nonumber \\
& \times & \phi_A(x_1,k_T^2;\mathbf{x}_\perp)
               \phi_B(x_2,(p_T-k_T)^2;\mathbf{x}_\perp), 
\label{eq:ktfac}
\end{eqnarray}
where
$x_{1,2}=p_T\exp(\pm y)/\sqrt{s}$, and $y$ and
$p_T$ are a rapidity
and a transverse momentum of a produced gluon.
Running coupling constant $\alpha_s$ is evaluated at the scale
$Q^2 = \max(k^2_T,(p_T-k_T)^2)$.
The unintegrated gluon distribution $\phi$ is related to
the gluon density of a nucleus by
\begin{equation}
   xG_A(x,Q^2) = \int^{Q^2} d^2k_T d^2x_\perp \phi_A(x,k^2_T;\mathbf{x}_\perp).
\end{equation}
Motivated by the KLN approach, we use a simplified
assumption about the unintegrated gluon distribution
function:
\begin{equation}
\label{eq:uninteg}
\phi_A(x,k^2_T;\mathbf{x}_\perp)\,\,
  =\left\{\begin{array}{l}
   \frac{\kappa C_F}{2\pi^3\alpha_s(Q^2_s)}\frac{Q_s^2}{Q_s^2+\Lambda^2}, \,                      \quad  k_T\,\leq\,Q_s, \\
   \frac{\kappa C_F}{2\pi^3\alpha_s(Q^2_s)}\, \frac{Q^2_s}{k^2_T+\Lambda^2},
              \quad k_T\,>\,Q_s,
\end{array}
\right.
\end{equation}
where $C_F=(N_c^2-1)/(2N_c)$. We introduce a small regulator
$\Lambda=0.2$ GeV/$c$ in order to
have a smooth distribution in the forward rapidity region
$|y|>4.5$ at RHIC.
Other regions are not affected by introducing a small regulator.
The above distribution depends on the transverse
coordinate $\mathbf{x}_\perp$ through $Q_s^2$.
Saturation momentum of a nucleus $A$
in $A+B$ collisions is obtained by solving the following
implicit equation with respect to $Q_s$
at fixed $x$ and $\mathbf{x}_{\perp}$
\begin{equation}
\label{eq:saturation}
 Q^2_s(x, \mathbf{x}_{\perp}) = \frac{2\pi^2}{C_F}
\alpha_s(Q^2_s)xG(x,Q^2_s)
              \rho^A_{\mathrm{part}}(\mathbf{x}_{\perp}),
\end{equation}
where 
\begin{equation}
   \rho^A_{\mathrm{part}}(\mathbf{x}_{\perp}) = T_A(\mathbf{x}_{\perp})
 \left\{1 - [1-
\sigma^{\mathrm{in}}_{NN}T_B(\mathbf{x}_{\perp})/B ]^B
          \right\}.
\end{equation}
The number of participants is given by
\begin{equation}
   N_{\mathrm{part}} = \int d^2x_{\perp}
           ( \rho^A_{\mathrm{part}}(\mathbf{x}_{\perp})
           +\rho^B_{\mathrm{part}}(\mathbf{x}_{\perp}) ).
\end{equation}
We take a simple perturbative form
for the gluon distribution in a nucleon
\begin{equation}
\label{eq:xG}
  xG(x,Q^2) = K\ln\left( \frac{Q^2 + \Lambda^2}{\Lambda_{\mathrm{QCD}}^2}\right)x^{-\lambda} (1-x)^n
\end{equation}
where
$\Lambda = \Lambda_{\mathrm{QCD}}=0.2$ GeV.
$K$ is used to control
saturation scale in Eq.~(\ref{eq:saturation})~\cite{BKW}.
We choose $K=0.7$  for $\lambda=0.2$
so that
the average saturation momentum in the transverse plane yields
$\langle Q_s^2(x=0.01)\rangle \sim 2.0$ GeV$^2/c^2$
in Au+Au collisions at impact parameter $b=0$ at RHIC.
Similar to the KLN approach, $x^{-\lambda}$ dependence of
the saturation scale
is motivated
by the Golec-Biernat--W\"usthoff model~\cite{GBW}.
The factor $(1-x)^n$ shows
that gluon density becomes small at $x\to1$.
$n$ usually depends on the scale $Q^2$.
Here we take $n=4$ as in the KLN approach~\cite{KLN}.

We obtain the rapidity distribution for produced gluons
at each transverse point $\mathbf{x}_\perp$
by performing the integral of Eq.~(\ref{eq:ktfac})
numerically.
The transverse energy distribution
$dE_T/dy$ is also obtained
 by weighting the transverse momentum of gluons
$p_T$ in the integration with respect to $p_T$
in Eq.~(\ref{eq:ktfac}).
We cut off the integral range of $p_T$ 
in Eq.~(\ref{eq:ktfac}) since
only the low $p_T$ partons are assumed to
reach the local thermal equilibrium.
We set the cut-off momentum as $p_{T,\mathrm{cut}}=3$
GeV/$c$ which corresponds
roughly to the maximum saturation scale at $x=0.01$
at the origin $\mathbf{x}_{\perp} = \mathbf{0}$
in central Au+Au collisions at RHIC.

In order to obtain initial conditions for hydrodynamics,
one needs a non-equilibrium description
for the collisions of heavy nuclei.
Leaving this problem for the future work,
we assume that the system of gluons initially produced
from the CGC
reaches a kinematically and chemically
equilibrated state at a short time scale.

There are two ways
to provide initial conditions from the CGC,
i.e., matching of number density  (IC-$n$)
and matching of energy density (IC-$e$).
Here, we employ the former prescription, i.e., IC-$n$. 
For the result from IC-$e$, see Ref.~\cite{HiranoNara6}.
Assuming Bjorken's ansatz $y=\eta_{\mathrm{s}}$ \cite{BJ}
where $\eta_{\mathrm{s}}$
is the space-time rapidity $\eta_{\mathrm{s}}=(1/2)\ln(x^+/x^-)$, 
we obtain the number density
for gluons at a space-time point
$(\tau_0, \mathbf{x}_\perp, \eta_{\mathrm{s}}) \equiv(\tau_0, \vec{x})$
from Eq.~(\ref{eq:ktfac})
\begin{eqnarray}
\label{eq:g_density}
n_g(\tau_0,\vec{x}) = \frac{dN_g}{\tau_0 d\eta_{\mathrm{s}} d^2x_\perp}.
\end{eqnarray}
Number density $n$ for the massless free parton system
can be written:
\begin{eqnarray}
  n &=& \left( \frac{3}{4}d_q + d_g \right) \frac{\zeta(3)}{\pi^2}T^3,
\end{eqnarray}
where $d_q=2N_cN_sN_f=36$, $d_g=2(N_c^2-1)=16$
and $\zeta(3)=1.20206$.
Thus,
we obtain the initial temperature field
in the three-dimensional (3D) space
from the number density (IC-$n$):
\begin{eqnarray}
T(\tau_0, \vec{x}) = \left\{\frac{\pi^2 n(\tau_0, \vec{x})}
{43\zeta (3)} \right\}^{1/3}.
\end{eqnarray}
When the temperature at ($\tau_0$, $\vec{x}$)
is below the critical temperature $T_{\mathrm{c}}=170$ MeV,
the all thermodynamic variables in the cell are set to zero.

\begin{figure}[ht]
\begin{center}
\includegraphics[width=0.45\textwidth]{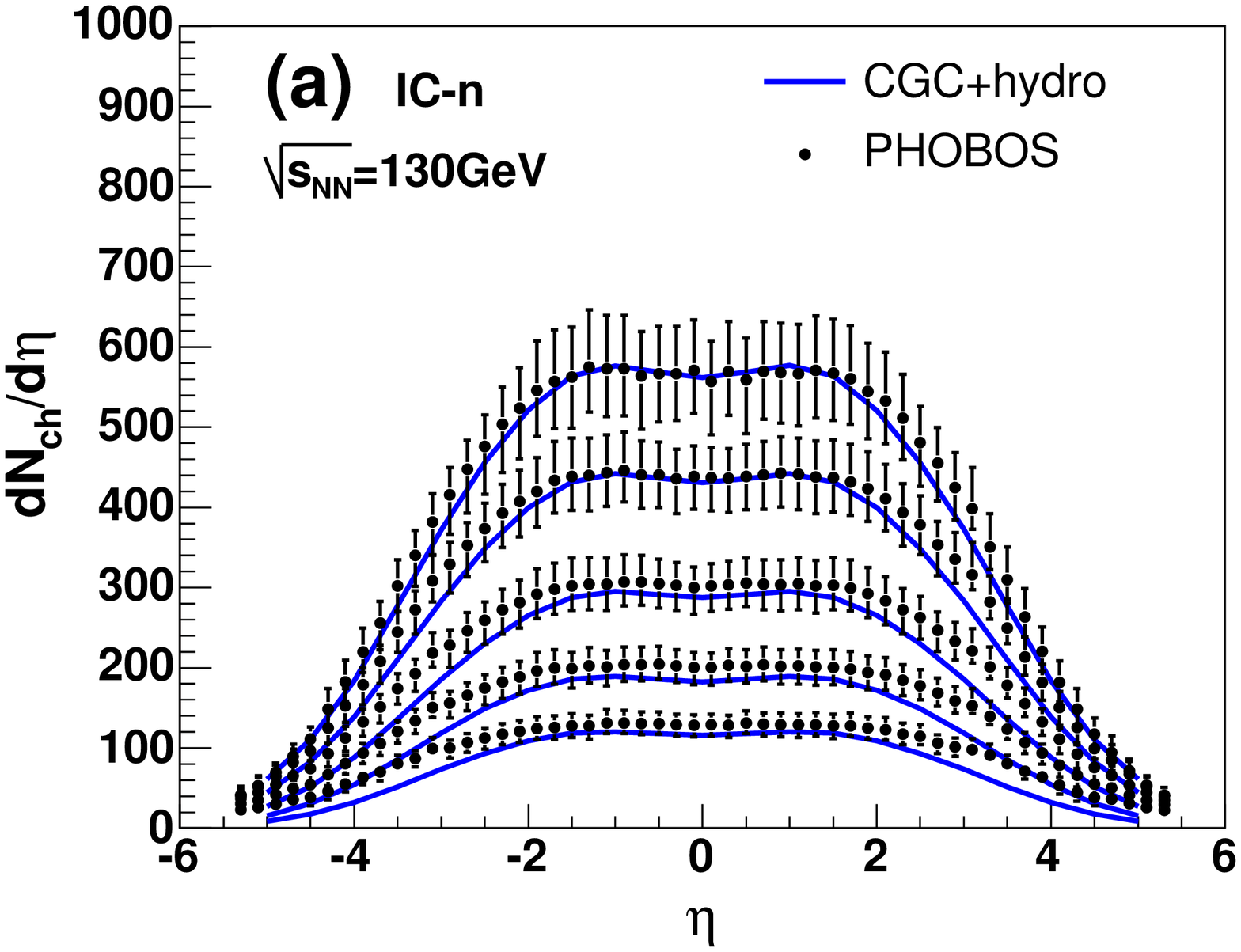}
\includegraphics[width=0.45\textwidth]{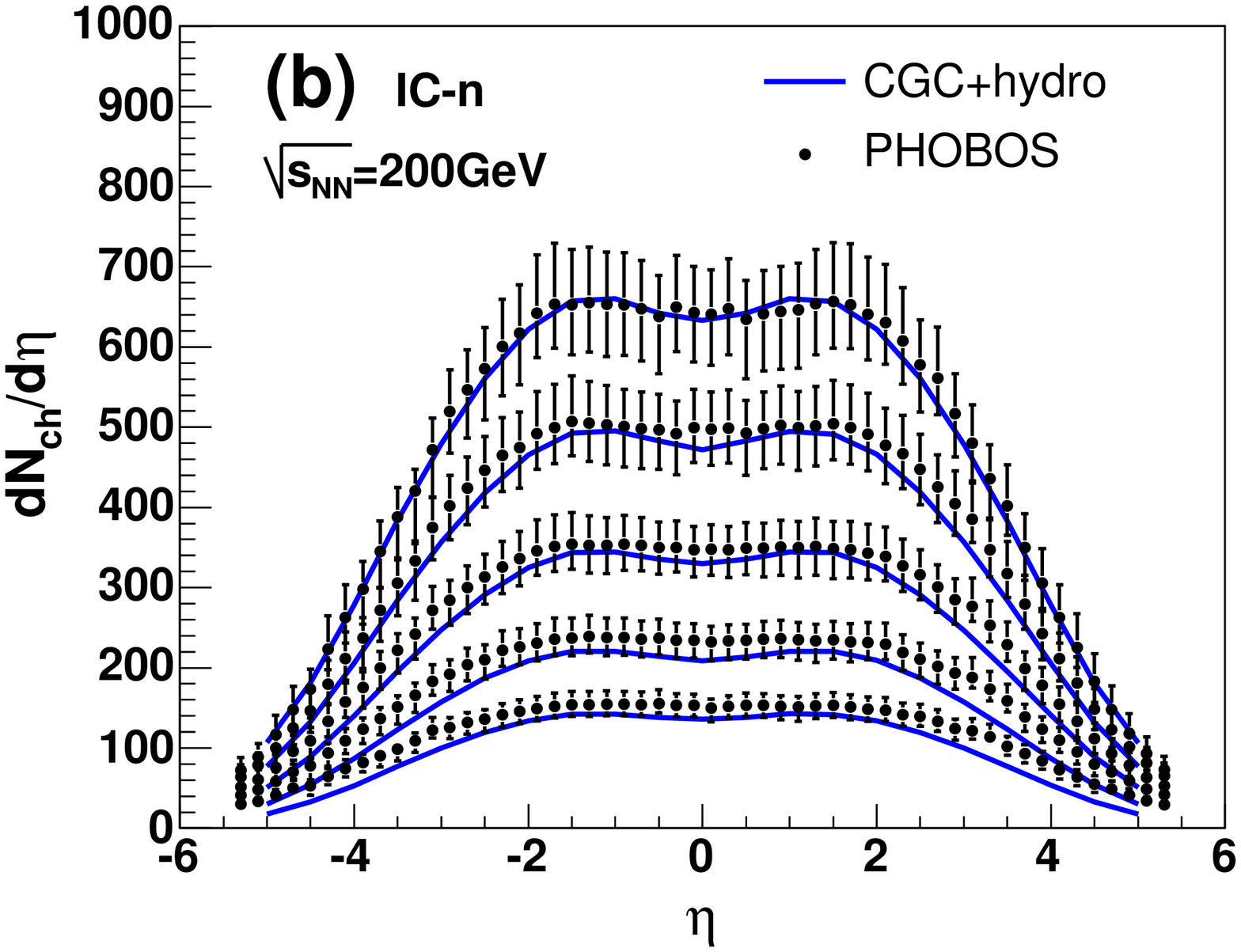}
\end{center}
\caption{
Pseudorapidity distributions of charged hadrons
 in Au + Au  collisions at $\sqrt{s_{NN}}$= (a) 130 and (b) 200 GeV
are compared to the PHOBOS data~\cite{Back:2002wb}.
Figure taken from Ref.~\cite{HiranoNara6}
}
\label{fig:dndeta200}
\end{figure}

\subsection{Results}

In Fig.~\ref{fig:dndeta200}, pseudorapidity distributions
of charged hadrons in Au + Au collisions at both $\sqrt{s_{NN}}=130$
and 200 GeV are compared with the PHOBOS data~\cite{Back:2002wb}.
Impact parameters in each panel
are, from top to bottom,
2.4, 4.5, 6.3, 7.9, and 9.1 fm
(2.5, 4.4, 6.4, 7.9, and 9.1 fm)
for $\sqrt{s_{NN}} =$ 200 (130) GeV.
These impact parameters
are evaluated from
the average number of participants
at each centrality
estimated by PHOBOS~\cite{Back:2002wb}.
It should be also emphasized that it is not easy to parametrize
such initial conditions which reproduce the data with the same
quality as the CGC initial conditions presented here.

\begin{figure}[t]
\begin{center}
\includegraphics[width=0.45\textwidth]{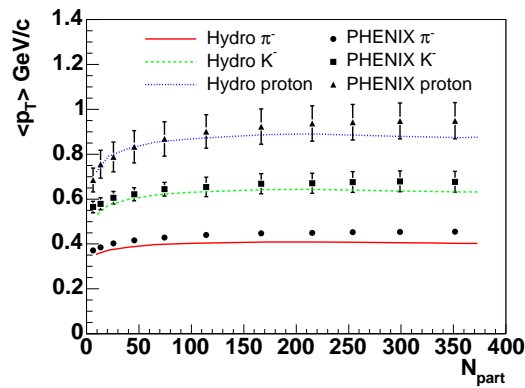}
\end{center}
\caption{
Mean transverse momenta for pions, kaons and protons
as a function of $N_{\mathrm{part}}$ \cite{phenix:pi}.
Here contribution only from hydrodynamic components is taken into account.
$T^{\mathrm{th}}=100$ MeV is used for all centralities.
Figure taken from Ref.~\cite{HiranoNara6}
}
\label{fig:meanpt}
\end{figure}

Mean transverse momenta $\langle p_T \rangle$
for pions, kaons and protons
as a function of $N_\mathrm{part}$ 
are compared with the PHENIX data \cite{phenix:pi}
in Fig.~\ref{fig:meanpt}.
Although our results are slightly smaller than the data
in central and semicentral regions, the overall trend is consistent
with the data.
For pions, the semihard spectrum starts to be comparable with the soft
spectrum around $p_T=1.5$-2.0 GeV/$c$.
We reproduce the $p_T$ spectra for pions by including contribution
from quenched jets.
So the deviation for pions can be filled by the semihard spectrum.
While semihard components for kaons and protons are very small
in low and intermediate $p_T$ regions.
A little more radial flow
is needed to gain the mean transverse momentum
in central collisions.

\begin{figure}[ht]
\begin{center}
\includegraphics[width=0.3\textwidth]{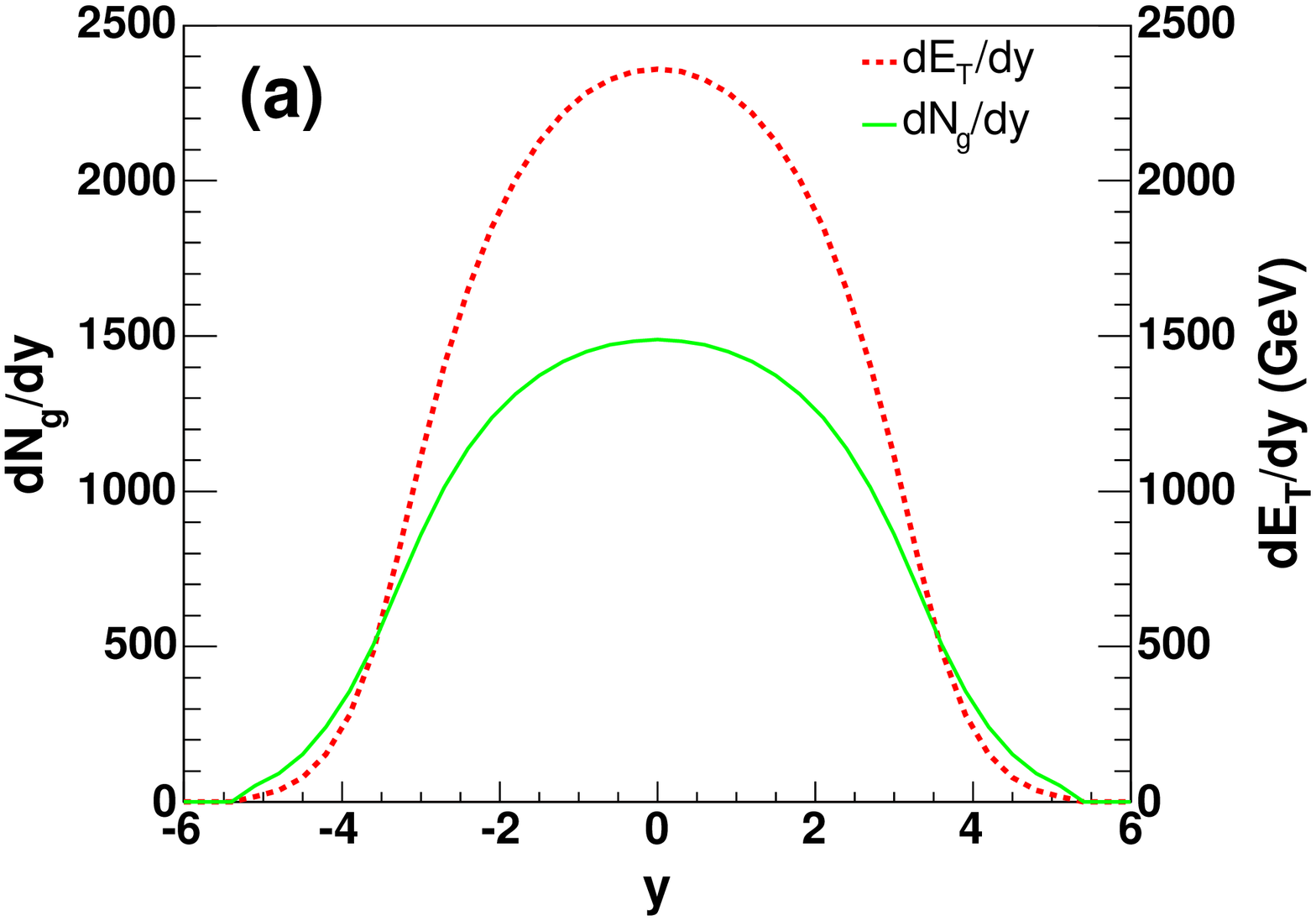}
\includegraphics[width=0.3\textwidth]{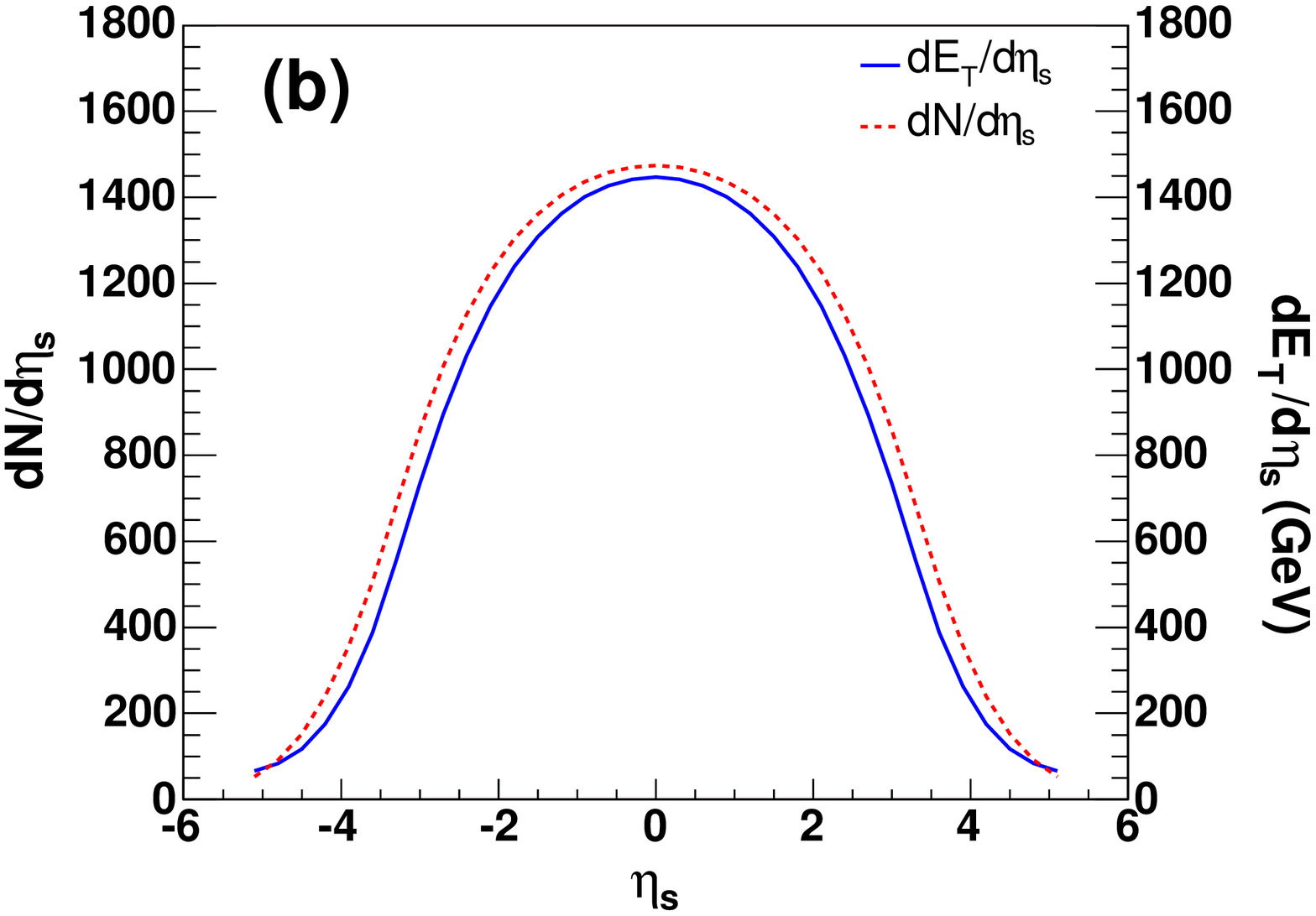}
\includegraphics[width=0.3\textwidth]{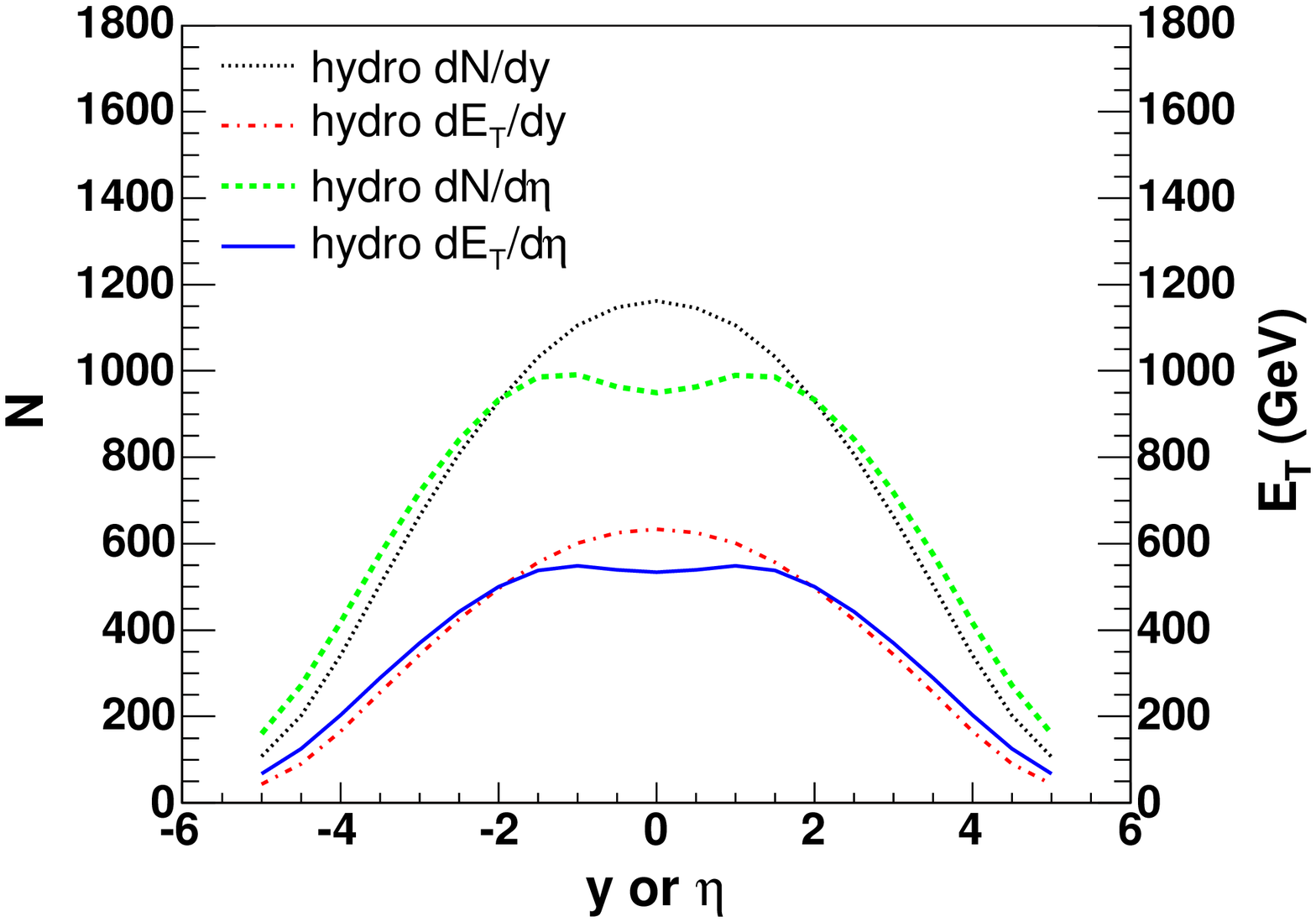}
\end{center}
\caption{
(Pseudo)rapidity distribution of (a) gluons produced from CGC collisions,
(b) partons at initial time, and (c) final total hadrons
 in Au + Au  collisions at $\sqrt{s_{NN}}=$ 200 GeV.
}
\label{fig:dndy}
\end{figure}

Both the KLN approach and the hydrodynamic model with a
CGC initial condition reproduce the centrality and rapidity
dependences of particle distributions.
Then, what is the role of hydrodynamic evolution
in comparison with the KLN approach?
The main difference between these two approaches
is whether final state interactions are taken into account.
The initial transverse energy per particle is estimated to be
$E_T/N_g \sim 1.6$ GeV at $y=0$.
It becomes $E_T/N_g=1.0$ GeV after assuming a thermal state.
The effect of the hydrodynamic afterburner is to reduce
the transverse energy per particle
due to $pdV$ work.
We find that $(dE_T/dy)/(dN/dy)|_{y=0} = 0.54$ GeV
and $(dE_T/d\eta)/(dN/d\eta)|_{\eta=0}=0.56$ GeV.

\subsection{Summary and Outlook}

Our approach is a first step toward a unified dynamical
modeling of relativistic heavy ion collisions.

The main goal has been developing a consistent
dynamical framework
for the space-time evolution of both bulk matter and hard
jets.
In order to accomplish this,
much remains are to be done.
\begin{itemize}

\item More sophisticated wave function
should be used to calculate the production of initial
gluons. At midrapidity, results from the classical Yang-Mills
equation can be used. In forward rapidity region
at the RHIC energy
or even at midrapidity at the LHC energy,
one needs a quantum evolution.

\item As discussed in the Sec.~1,
the hadron phase should be described by
a hadronic transport model. This will improve
$v_2$ in forward rapidity regions in the hydrodynamic result.

\item EoS from the recent lattice QCD simulations should be used
in the hydrodynamic simulations.

\item Chemical non-equilibrium process will be taken into account
by solving the rate equation for the number of quarks and gluons
together with the hydrodynamic equations \cite{biro}.
The amount of energy loss of partons depends on
a species of both probe parton and medium parton.
So the correct chemical composition in the QGP phase
will lead to reproduction of nuclear modification factors
without any free parameters.

\end{itemize}

\ack
One of the authors (T.H.) would like to thank the
organizers of this workshop for invitation.
The work by T.H. was supported by RIKEN.

\end{document}